\documentclass[conference]{IEEEtran}
\usepackage{graphicx}
\usepackage[table,xcdraw]{xcolor}
\usepackage{subfig}
\usepackage[font=small,labelfont=bf,tableposition=top]{caption}
\usepackage{enumerate}
\usepackage[numbers]{natbib}
\usepackage{hyperref}
\begin{document}
	
	\title{Most central or least central? How much modeling decisions influence a node's centrality ranking in multiplex networks}

	\author{\IEEEauthorblockN{Sude Tavassoli}
		\IEEEauthorblockA{Gottlieb-Daimler-Str. 48-668\\
			Computer Science Department\\Kaiserslautern University of Technology\\67663 Kaiserslautern, Germany\\
			Telephone: +49-631-2053340\\
			Email: tavassoli@cs.uni-kl.de}
		\and
		\IEEEauthorblockN{Katharina A. Zweig}
		\IEEEauthorblockA{Gottlieb-Daimler-Str. 48-672\\
			Computer Science Department\\Kaiserslautern University of Technology\\67663 Kaiserslautern, Germany\\
			Telephone: +49-631-2053346\\
			Email: zweig@cs.uni-kl.de}}
	
	\maketitle
	
	\begin{abstract}
	To understand a node's centrality in a multiplex network, its centrality values in all the layers of the network can be aggregated. This requires a normalization of the values, to allow their meaningful comparison and aggregation over networks with different sizes and orders. The concrete choices of such preprocessing steps like normalization and aggregation are almost never discussed in network analytic papers.
In this paper, we show that even sticking to the most simple centrality index (the degree) but using different, classic choices of normalization and aggregation strategies, can turn a node from being among the most central to being among the least central. 
We present our results by using an aggregation operator which scales between different, classic aggregation strategies based on three multiplex networks. We also introduce a new visualization and characterization of a node's sensitivity to the choice of a normalization and aggregation strategy in multiplex networks. 
The observed high sensitivity of single nodes to the specific choice of aggregation and normalization strategies is of strong importance, especially for all kinds of intelligence-analytic software as it questions the interpretations of the findings.
	\end{abstract}

	\IEEEpeerreviewmaketitle
	
\section{Introduction}
	Classical Centrality indices have been numerously used and extended to quantify the importance of nodes in a network with respect to various network processes~\cite{Borgatti1994,Freeman1979,Koschuetzki2005,Borgatti2005}. It was soon noticed that a comparison of centrality index values of nodes in different networks requires a careful normalization: a degree of $100$ might be considered more important than a degree of $200$, if the first occurs in a network of order $150$ and the second occurs in a network of order $500$~\cite{Freeman1979}. Classic normalization strategies, e.g., divide the degree by the order of the network, by the maximal observed value, or by first subtracting the minimal observed degree and then by dividing by the difference between the maximal and minimal observed degrees. 

Computing the centrality of a node is more complicated when a multiplex network comprised of multiple layers is under analysis. The rapidly growing number of studies of multiplex networks in network science per se indicates that the analysis of complex systems needs more comprehensive network models and frameworks~\cite{Kivela2014}. In such networks (also called multilayer networks), each layer itself is a network and the nodes are connected to each other with respect to different types of interaction or relationships~\cite{Kivela2014}. While some authors allow edges between nodes from different layers~\cite{De2013,De2015,De2013anatomy,De2013math}, here we restrict ourselves to multiplex networks comprised of multiple, non-interconnected layers with at least some nodes common to all layers. 

If now an evaluation of the centrality of a node is needed for multiplex networks, different strategies have been proposed. Given a centrality index and a multiplex network, most authors suggest to use a vector of centrality index values, where each entry corresponds to the node's value in one layer~\cite{Battiston2014,Boccaletti2014}. This does not easily allow for a single ranking of all nodes. Some authors have thus suggested to either use the sum or average of all centrality index values~\cite{Boccaletti2014}; other strategies could be to rank by the maximum value achieved or by the minimum value obtained, to either stress the most successful ranking within one layer or the least successful and thus minimal centrality over all layers.
	
In most network analytic studies comparing or aggregating centrality indices, neither the choice of the normalization nor the aggregation strategy is discussed---sometimes it is not even reproducibly described. The main reason for this seems to be that these choices seem to be so inconsequential. However, this question has not been addressed so far.

	Thus, in this paper, we show that even the most simple centrality index, the {\it degree} of a node, when normalized and aggregated in various ways is sensitive to the corresponding choices, to various degrees. We show this on three multiplex network data sets from three very different complex systems: A data set representing a part of the European air transportation network, a set of Twitter related networks between people interested in the Higgs-Boson research, and a data set describing different relationships between employees of a law firm. We also introduce a visual method how to categorize nodes by their sensitivity to either the choice of aggregation or normalization strategy, to both, or to none. This visualization helps to decide whether the choice of a normalization and aggregation strategy needs to be defended or not.
	
	The remainder of this paper is organized as follows: Section~\ref{def} presents definitions, data sets, normalization and aggregation methods. Section~\ref{Experimentalresults} presents results obtained on three different multiplex networks and Section~\ref{summary} summarizes the paper.	
\section{\label{def}Definitions, Data sets, Normalization and Aggregation Strategies}

 We define a multiplex network as a network with $|\mathcal L|$ layers $\mathcal L=\{L_1,L_2,\cdots,L_{|L|}\}$ where each layer $l_i$ is a simple graph comprised of a node set $V_i$ and an edge set $E_i\subseteq V_i\times V_i$, with $n_i:=|V_i|$ nodes and $m_i:=|E_i|$ edges. Each edge set $E_i$ represents a specific type of interaction; The set of nodes common between layers $l_1$ to $l_{|L|}$ is denoted by $V^*= \bigcap_{i=1}^{|L|} V_i$. 
The degree $deg_i(v)$ of any node $v$ is defined as the number of edges connected to the node $v$ in layer $L_i$.

 \subsection{Datasets}
 \label{datasets}
 The first multiplex network data set is a subset of the \textbf{European Airlines Network}, which encompasses $37$ layers of airlines and which was compiled by Cardillo et al.~\cite{Cardillo2013}. The airports are represented as nodes and two nodes are connected if the corresponding airline offers a flight between the corresponding cities. Since the $37$ layers do not share common nodes, we use the the following two subsets of airlines: Air Berlin, Easyjet, Lufthansa, and Ryan air. The details of these four layers are listed in Table~\ref{tab:threetables_Airline}; the four air lines share $9$ airports. We also use a further subset by removing Lufthansa; it results in $20$ shared airports among Air Berlin, Easyjet, and Ryan Air.

  The second dataset is a \textbf{Twitter related network} of users interested in research on the Higgs Boson particle, compiled by De Domenico et al.~\cite{De2013anatomy}. The different layers represent four directed networks of users {\it mentioning} other users in their tweets, users {\it replying} to the tweets of other users, users {\it re-tweeting} the tweets os other users, and the social network of based on the users {\it following} other users~\cite{De2013anatomy}. We restrict our analysis to the biggest, strongly connected component of each of these four layers; these share $|V^*|=127$ nodes. The details of the four layers of this dataset are shown in Table~\ref{tab:threetables_Twitter}.

  The third multiplex network represents different relationships between the employees of a \textbf{Law firm dataset}, compiled by Lazega (2001)~\cite{Lazega2001} in a study of how $71$ attorneys of a law firm communicate in terms of seeking advice, co-Working, and having a friendship outside the firm. The three relationships are represented in three layers of a multiplex network. In this dataset, we have $|V^*|=|V_1|=|V_2|= |V_3|=71$ common nodes. Likewise the other datasets, the details are listed in Table~\ref{tab:threetables_Lawfirm}.

  \begin{table}[t]		
 \caption{Properties of all the layers of the three multiplex network data sets; $V^*$ is defined as the set of nodes common to all layers of the respective dataset.} \label{tab:threetables}
\centering
\subfloat[European Airlines Network dataset. The number of common nodes in the subset of Lufthansa, Air Berlin, Ryan Air and Easyjet is $|V^*|=9$.]{\label{tab:threetables_Airline}
\centering
\scriptsize
\label{tab:Tab5}
\begin{tabular}{|l||l||l||l||l|}
	\hline
     Properties          & Air-Berlin & Easyjet & Lufthansa & Ryanair \\\hline
$|V_i|$      & 75         & 99      & 106       & 128     \\
$|E_i|$      & 239        & 347     & 244       & 601     \\
$\max_{v\in V_i} \{deg(v)\}$ & 37         & 67      & 78        & 85      \\
$\max_{v\in V^*} \{deg(v)\}$ & 26         & 17      & 5        & 28      \\
$\min_{v\in V_i} \{deg(v)\}$ & 1          & 1       & 1         & 1       \\
$\min_{v\in V^*} \{deg(v)\}$ & 1          & 2       & 1        & 5       \\
\hline
\end{tabular}
}
\\
\centering
\subfloat[Twitter Network dataset. The number of common nodes, based on the corresponding biggest, strongly connected component  in each of the four layers is $|V^*|=127$.]{\label{tab:threetables_Twitter}
\centering
\scriptsize
\label{tab:Tab7}
\begin{tabular}{|l||l||l||l||l|}
	\hline
     Properties          & Mention & Reply & Retweet & SocialNetwork \\\hline
$|V_i|$     & 1801         &  322      & 984       & 360210     \\
$|E_i|$     & 7069        & 708     & 3850       & 14102605     \\
$\max_{v\in V_i} \{deg(v)\}$ & 466         & 45      & 212        & 44611      \\
$\max_{v\in V^*} \{deg(v)\}$ & 141         & 45      & 101        & 33664      \\
$\min_{v\in V_i} \{deg(v)\}$ & 2          & 2       & 2        & 2       \\
$\min_{v\in V^*} \{deg(v)\}$ & 2          & 2       & 2        & 27       \\
\hline
\end{tabular}
}
\\
\centering
\subfloat[Law firm dataset. The three following layers share the same $71$ nodes, i.e., $V_i=V^*$ for all layers.]{\label{tab:threetables_Lawfirm}
\centering
\scriptsize
\label{tab:Tab8}
\begin{tabular}{|p{4.81cm}||p{0.53cm}||p{0.82cm}||p{0.5cm}|}
	\hline
     Properties          & Advice & Coworker & Friend  \\\hline
$|V_i|$      & 71         &  71      & 71         \\
$|E_i|$      & 717        &   726  & 399        \\
$\max_{v\in V_i} \{deg(v)\}=\max_{v\in V^*} \{deg(v)\}$ & 46         & 45      & 28  \\

$\min_{v\in V_i} \{deg(v)\}= \min_{v\in V^*} \{deg(v)\}$ & 3          & 7       & 1           \\
\hline
\end{tabular}

}	
  \end{table}
 
As can be seen in the tables, the layers of the first two data sets vary in their order, i.e., the number of nodes. In the Airline data set, the order varies between $75$ and $128$ nodes, in the Twitter data set, the variance is even larger between $322$ nodes in the ``reply'' layer and $360,210$ nodes in the social network based on the ``followers/followees'' relationship. When we want to rank the nodes that are common to all layers, it is obvious that any meaningful aggregation of a degree centrality value can only be obtained by normalizing the values, as described in the following.

\subsection{Different Normalization Methods}
\label{NormMethods}
The result of the normalization and aggregation is a ranking of all common nodes in $V^*$, from position $1$ (among them) to position $|V^*|$. Here, we describe how any node's degree $deg_i(v)$ is normalized from which then only the nodes in $V^*$ are selected. 

\textbf{NormMethod 1}, for layer $L_i$ takes $deg_i(v)$ for all $v \in V^*$ and normalizes it with the minimum and maximum values in the set of common nodes. This results in a vector of normalized indices of $[0,1]$ for layer $L_i$.

\[\mathcal{C}_1(v,i)= \frac{deg_i(v)-min\{deg_i(v)|v\in V^*\}}{max\{deg_i(v)|v\in V^*\}-min\{deg_i(v)|v\in V^*\}}\]

I.e., for the airlines network, all nodes in layer $1$ (Air Berlin) are normalized by subtracting $1$ and dividing by $(26-1)$; layer $2$ (Easyjet) subtracts $2$ and divides by $(17-2)$; layer $3$ (Lufthansa) subtracts $1$ and divides by $(5-1)$; layer $4$ (Ryanair) subtracts $5$ and divides by $(28-5)$. 

\textbf{NormMethod 2} is a commonly used normalization in many studies. It is similar to the last method but the normalization is done using the minimum and maximum values in the set of all nodes ($V_i$) in layer $L_i$. Thus, among the nodes in $V^*$ which are ultimately ranked, there might or might not be a node with a normalized value of $0$ or $1$, depending on whether the node with minimal and maximal degree in $L_i$ are in $V^*$ or not.

I.e., for the airlines network, all nodes in layer $1$ (Air Berlin) are normalized by subtracting $1$ and dividing by $(37-1)$; layer $2$ (Easyjet) subtracts $1$ and divides by $(67-1)$; layer $3$ (Lufthansa) subtracts $1$ and divides by $(78-1)$; layer $4$ (Ryanair) subtracts $1$ and divides by $(85-5)$. 

\[\mathcal{C}_2(v,i)=\frac{deg_i(v)-min\{deg_i(v)|v\in V_i\}}{max\{deg_i(v)|v\in V_i\}-min\{deg_i(v)|v\in V_i\}}\]

\textbf{NormMethod 3} uses the results by \textit{NormMethod 2} and multiplies them with the fraction of the maximum degree in layer $L_i$ and the maximum degree among all nodes in all $|\mathcal L|$ layers. This results in a vector of indices of nodes ($v\in V_i$) between $[0,\frac{\max\{deg_i(v)|v\in V_i\}}{\max\{deg_i(v)|v\in \bigcup V_j, 1\leq i\leq |\mathcal L| \}}]$.

\[\mathcal{C}_3(v,i)=\mathcal{C}_2(v)\cdot \left(\frac{max\{deg_i(v)|v\in V_i\}}{max\{deg_i(v)|v\in \bigcup V_j, i\in [1, \ldots,|\mathcal L|]\}}\right)\]

I.e., for the airlines network, all nodes in layer $1$ (Air Berlin) take the values from NormMethod 2 and multiply by $37/85$; layer $2$ (Easyjet) multiplies by $67/85$; layer $3$ (Lufthansa) multiplies by $78/85$; layer $4$ (Ryanair) multiplies by $85/85)1$. 

Freeman already discovered that some networks are more {\it centralized} than others, i.e., that one node with a very high degree dominates the degrees of the others. In all of the above normalization methods, it might thus happen, that a node with the second-highest degree in a strongly centralized layer has a quite small normalized degree centrality. Nonetheless, it is the second-most central node in its layer. If the aggregation strategy wants to reward a node that is among the top most nodes in at least one layer but some layers are more centralized than others, any of the above normalization methods would fail. In a centralized layer, most nodes show a very small normalized degree and some show a large normalized degree. In the cumulative distribution of the normalized values, i.e., the percentage of nodes with at least normalized degree $x$ plotted against $x$, we see a sharp increase followed by a long tail until $1$ is reached. In a less centralized network, the cumulative distribution runs closer to the diagonal. 

\begin{figure}[]
       	\centering
       	\begin{tabular}{c}
       		\subfloat[European Airlines Network dataset.]{\includegraphics[height=0.3\textwidth,width=0.45\textwidth]{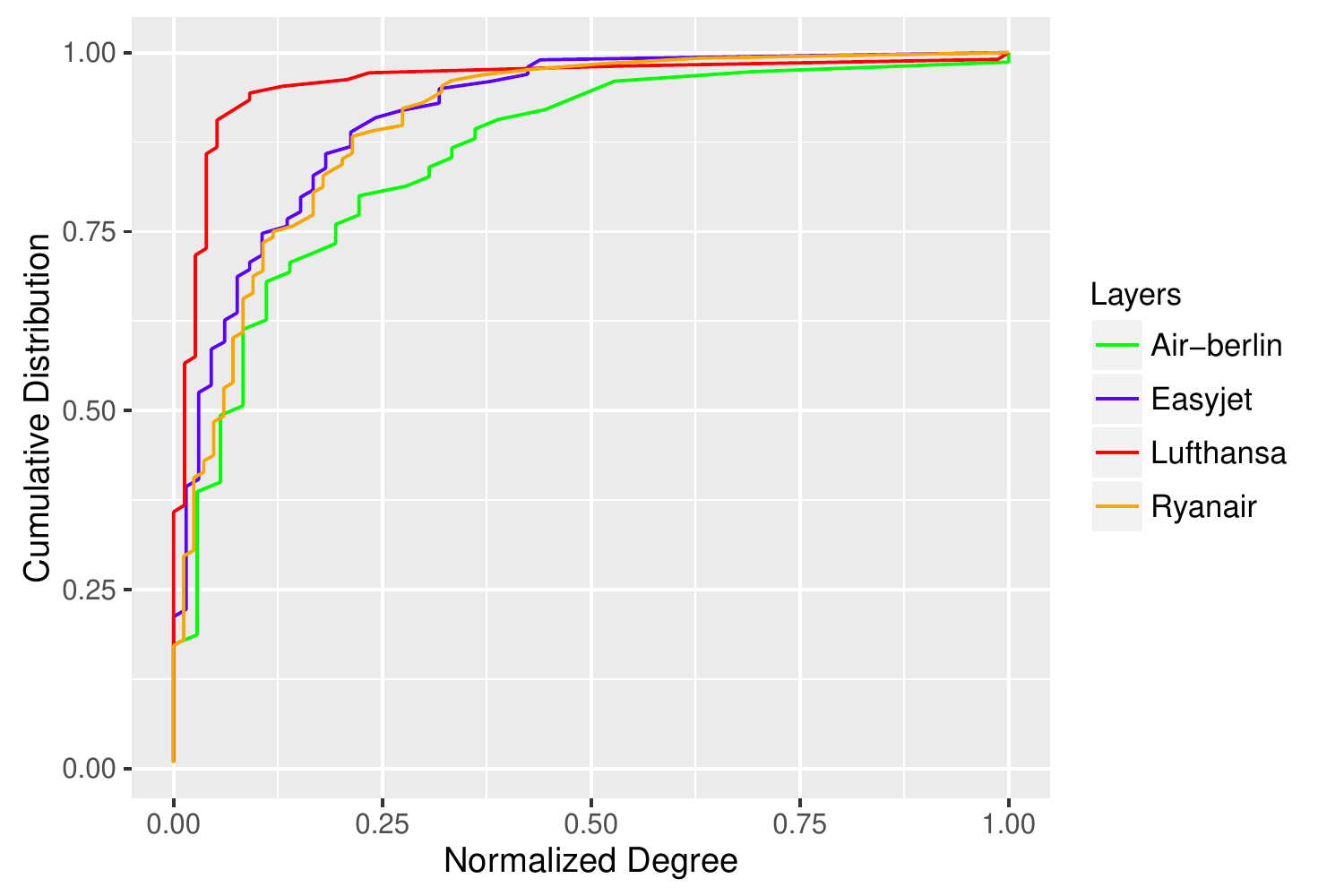}}\\	
       		\subfloat[Twitter Network dataset.]{\includegraphics[height=0.3\textwidth,width=0.45\textwidth]{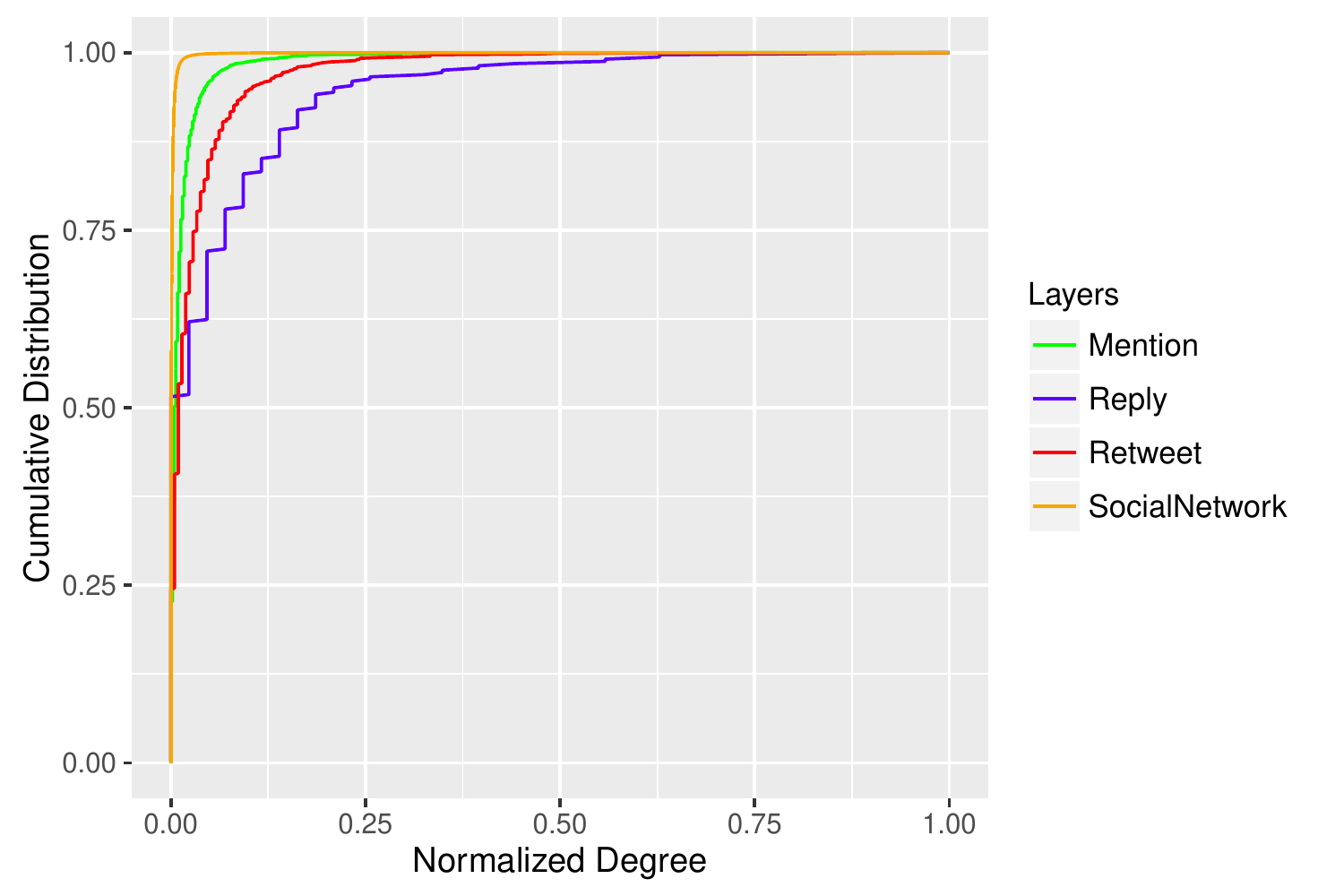}}\\
       		\subfloat[Law firm dataset. ]{\includegraphics[height=0.3\textwidth,width=0.45\textwidth]{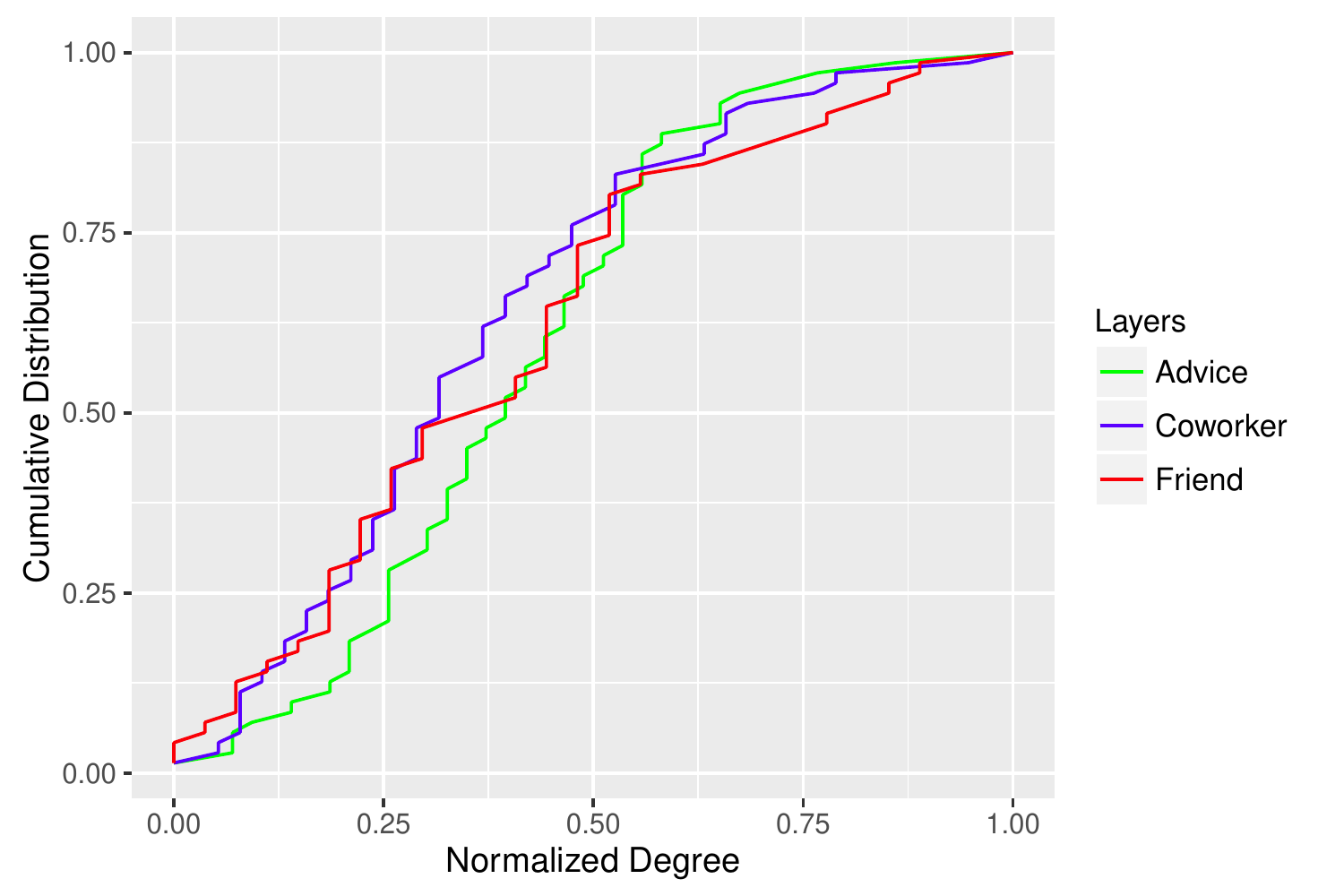}}
       	\end{tabular}
       	\caption{The cumulative distributions of the normalized degree (obtained using NormMethod 2) in all the layers of three multiplex network data sets. Some data sets contain more and less centralized layers at the same time which makes normalization difficult.}\label{fig:3cumdist}
       \end{figure}

For the three used network data sets, the results of NormMethod 2 are presented as a cumulative distribution (Figure~\ref{fig:3cumdist}). 
While the cumulative distributions of the normalized degree are similar for the layers of Easyjet and Ryan air, the other two layers show a different behavior. It becomes obvious that $90\%$ of the normalized degrees in the Lufthansa layer are smaller than $70\%$ of the normalized degrees in the Air Berlin layer. Thus, an aggregation strategy that wants to favor nodes that are among the most central nodes in at least one layer, would not be able to identify most central nodes in the Lufthansa layer as even medium central nodes in the Air Berlin layer would show a larger normalized degree centrality. 

Thus, the last normalization method which we will analyze looks at each node's position in rankings of the degree within each layer.


\textbf{NormMethod 4}  For each layer, we rank the nodes non-increasingly by their degree $deg_i(v)$ and obtain $r_i(v)$. This is then normalized by $n_i$. 
\[\mathcal{C}_4(v,i)=\frac{r_i(v)}{n_i}\]

\subsection{Different Aggregation Strategies}
\label{MEOWA}
In this paper, we consider the resulting normalized degree of $|L|$ layers as $|L|$ criteria and for the aggregation of these multiple criteria, we use \underline{M}aximum \underline{E}ntropy \underline{O}rdered \underline{W}eighted \underline{A}veraging (MEOWA) operator as described as follows:

The \textbf{MEOWA operator} has been proposed by Yager for aggregating multiple numerical values, that represent the satisfaction of different, possibly opposing criteria~\cite{Yager1998-2}. He stated that the aggregation of multiple criteria in a decision making problem for finding the best solution from a set of alternatives can be scaled between two extreme cases of pure $OR$ and pure $AND$. In the pure $OR$, the maximum value among the criteria is the result of aggregation and in the pure $AND$, the minimum value among the criteria is of interest. In the first case, this favors a solution which fulfills at least one criteria most satisfactorily, in the second case even the least satisfying criteria needs to be high. 
For the problem defined in this paper that means, we can either rank the nodes by their highest normalized degree in any layer or by their lowest. In the first case, the node needs to be most central in at least one layer, in the last case a node is only ranked high if its most peripheral position is still quite high. 

The MEOWA aggregation operator (denoted by $\lambda$) creates a single number based on the vector of a node's $|\mathcal L|$ normalized degrees as follows:

\[\lambda(C_x(v,1),C_x(v,2),\cdots,C_x(v,|\mathcal L|))=\sum_{j} w_j\ d_j(v)\]
where $D=(b_1, b_2, ..., b_{|\mathcal L|})$ is the non-increasingly sorted vector of the normalized degrees, and $w$ is a weight vector. The weight vector is obtained using the following function based on a parameter $\beta$~\cite{Yager1995}:
\[w_i=\frac{e^{\beta\frac{n-i}{n-1}}}{\sum_{j=1}^{n} e^{\beta\frac{n-j}{n-1}}}.\]
The produced weights are always between $[0,1]$ and the sum of the resulting weights is equal to $1$. For $\lim \beta=\infty$, the resulting weight vector is $(1,0,\ldots, 0)$. By multiplying this vector with the non-increasingly sorted normalized degrees of a node, its {\it highest} normalized degree is the result. When $\lim \beta=-\infty$, the weight vector is $(0,\ldots, 0,1)$ and this means the result is taking the minimum value among the normalized degrees of $v$. When $\beta=0$, then (for all $n$), the weight vector is simply given by $(1/n, 1/n, ..., 1/n)$. Thus, by multiplying this vector with the non-increasingly sorted normalized degrees of a node, a regular average is calculated. The first case follows the aggregation strategy that a node should be very central in {\bf at least one layer}, to be ranked high. The second case follows the aggregation strategy that a node should be very central in {\bf all} layers to be ranked high. The third case follows the aggregation strategy that a node should be {\bf on average} or {\it in many layers} very central, to be ranked high.
Any $\beta$-value between the extreme strategies of ``at least one'' and ``all layers'' can be interpreted as {\it scaling} and be described using a set of proportional linguistic quantifiers (\textit{a few, some, most, almost} introduced by Zadeh~\cite{Zadeh1965}~\cite{Zadeh1983-2}) -- we have discussed in our recent study~\cite{Tavassoli2015} how the most influential node in a network can be identified using a fuzzy operator that includes these linguistic quantifiers.

For historical reasons and most meaningful for binary criteria, Yager calls the first operator an $OR$-operator (only one layer determines the result) and the second operator an $AND$-operator (all layers determine the result). For any chosen $\beta$ and the corresponding operator, Yager defined an \textit{orness} measure denoted by $\Omega$~\cite{Yager1995}:

\[\Omega=\frac{1}{n-1}\sum_{i=1}^{n}(n-i)\frac{e^{\beta\frac{n-i}{n-1}}}{\sum_{j=1}^{n} e^{\beta\frac{n-j}{n-1}}}\]  

Yager shows that the entropy of the weight vector can be measured using the following equation (for more detail, see ~\cite{Yager1995}):
\[E(w)=-\sum_{i=1}^{n}w_i\cdot \ln w_i.\]

We use values between $[-20,20]$ for the parameter of $\beta$ which result in vectors with an $OR$ness of almost $0$ and $1$, respectively


In the following, we will show how sensitive a node's ranking is to the choices of $\beta$ and the normalization method.    

\section{Experimental Results}
\label{Experimentalresults}

\begin{figure}
 	\subfloat[Rankings of two airports chosen among $9$ common nodes in the four layers representing Air Berlin, Easyjet, Lufthansa, and Ryanair.]{\includegraphics[width=0.5\textwidth]{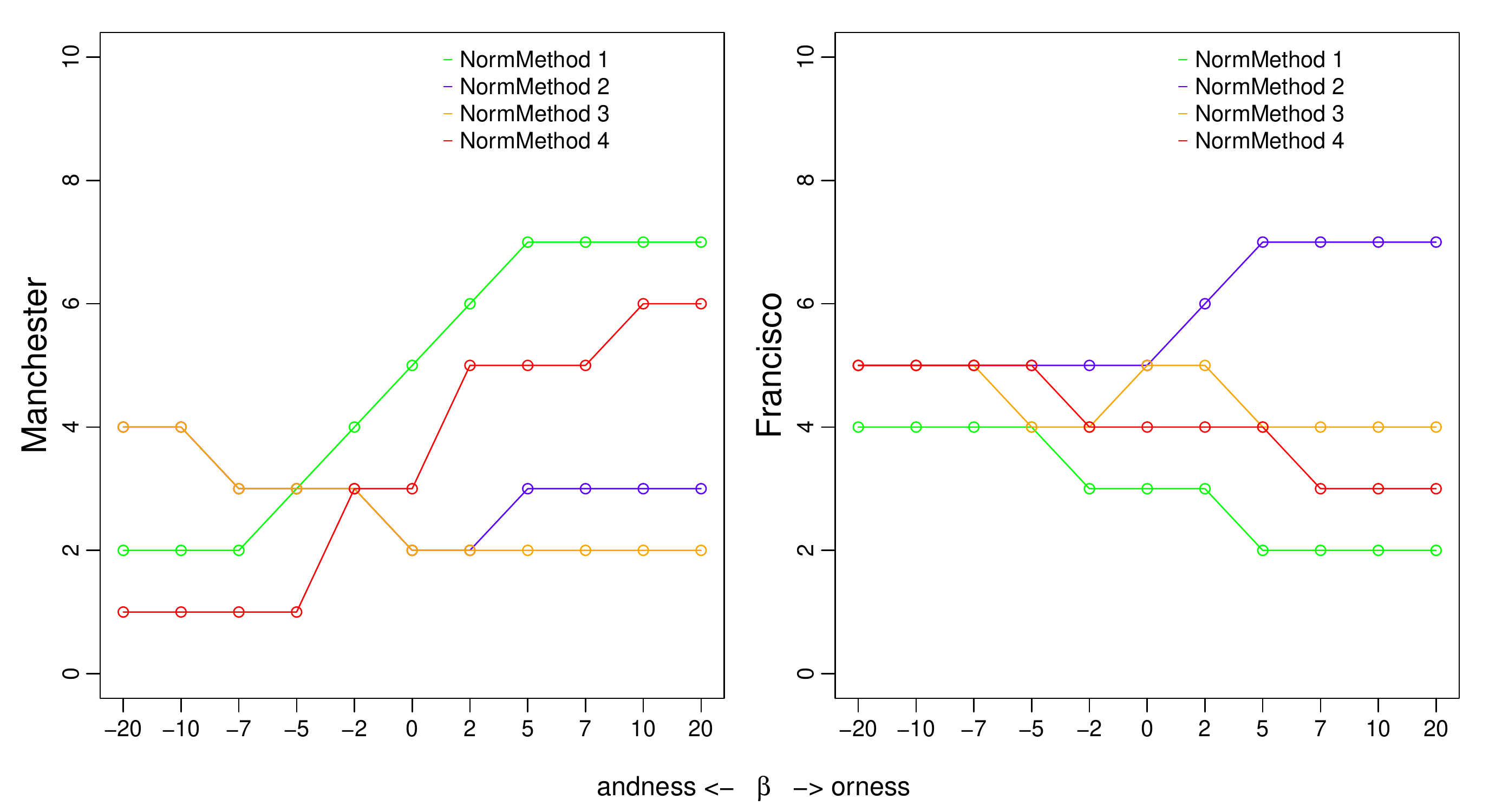}}\\
 	\subfloat[Rankings of $8$ airports selected from $20$ common nodes in the three layers representing Air Berlin, Easyjet, and Ryanair.]{\includegraphics[width=0.5\textwidth]{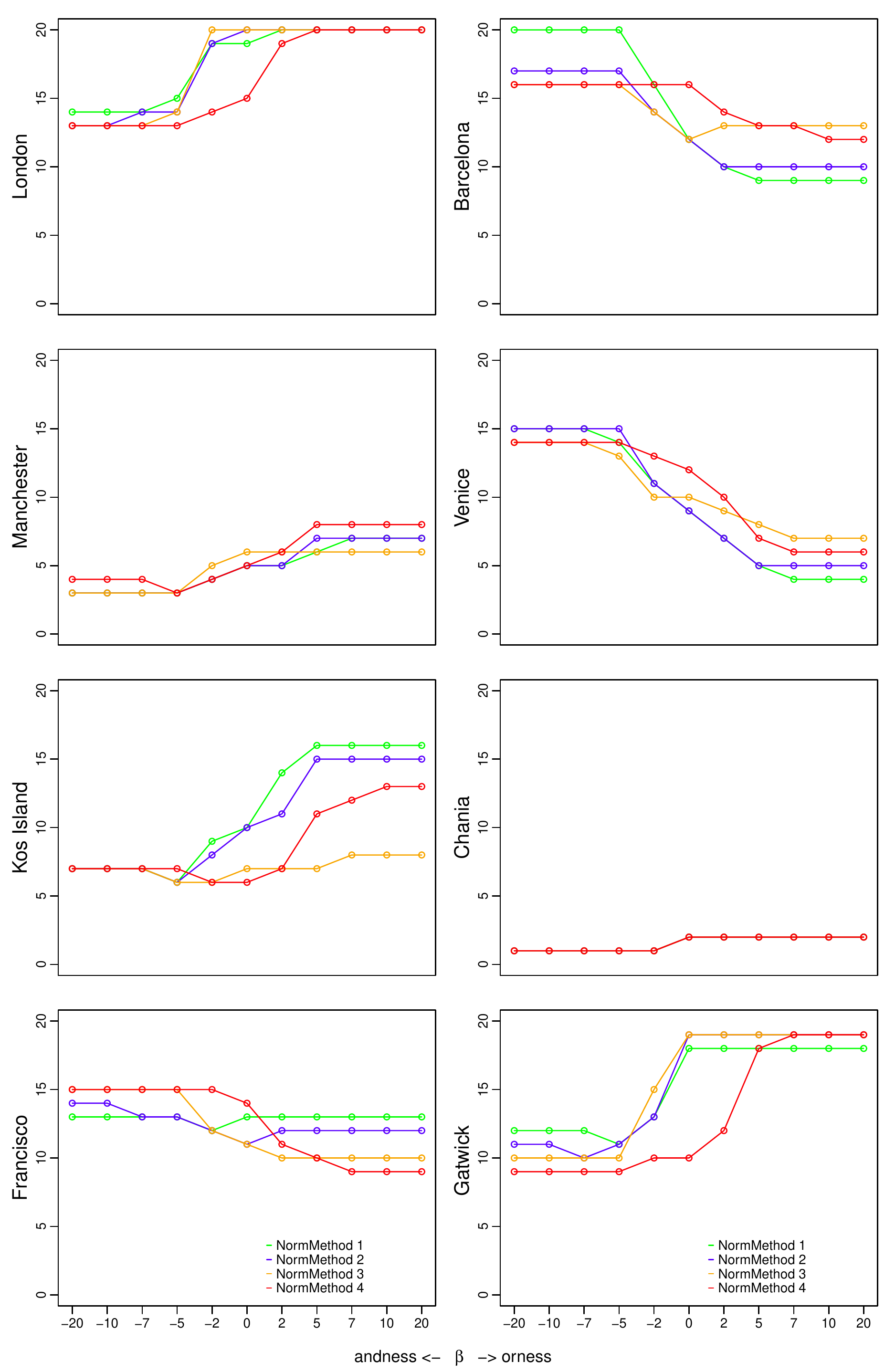}}
 	\caption{European airlines data set. Based on four different types of the normalized degree of the nodes in all layers, different aggregation strategies are obtained using the MEOWA operator steered by the $\beta$ parameter. The four colored curves show the ranking positions of the indicated node based on its normalized degrees, depending on $\beta$. \label{Fig1}} 
  \end{figure}

Figure~\ref{Fig1}.a) presents the resulting rankings of two airports in the airline networks (all four airlines), with one curve for each normalization method and plotted against the chosen $\beta$-value which represents the aggregation strategy. The rankings are relative rankings of the $9$ common nodes that exist in all four layers. Recall that a high $\beta$-value (high $OR$ness) favors nodes with a high normalized centrality degree in at least one layer, a $\beta$-value of $0$ (an $OR$ness of $0.5$) orders nodes by their average normalized centrality degree, and a low $\beta$-value (low $OR$ness = high $AND$ness) favors nodes with the highest minimal normalized degree.


Regarding all four layers, Manchester airport is an interesting case as shown in Figure.~\ref{Fig1}.a). 
Its actual degree in the four layers of airlines is: $(1, 12, 5, 5)$, respectively, while the maximal degree of all common nodes is: $(26,17,5,28)$ and the maximal degrees of all nodes in the respective layers is: $(37,67,78,85)$ (s. Table~\ref{tab:threetables_Airline}).
Now, follow the curve resulting from using NormMethod 1 along different $\beta$-values (i.e., different aggregation strategies): Manchester changes its position from rank $2$ among the $9$ common nodes to rank 7. NormMethod 1 normalizes with the maximal degrees of all common nodes in the same layer, and thus, in layer $3$ it gets a normalized degree of $1$; for $\beta=20$ and $n=4$, the weight vector multiplies the highest normalized degree by $0.999$ and the second highest by $0.001$.  There are two other nodes with the same maximal normalized degree of $1$ and a higher second-highest normalized degree than Manchester, thus, it gets position $7$ for $\beta=20$. Since its minimal normalized degree is very low, it is only at rank $2$ for $\beta=-20$. Fixing a normalization method, we see that Manchester is thus quite sensitive to the chosen aggregation strategy. To quantify this sensitivity, we define $\Delta agg$ as the maximum difference in ranking position fixing any of the normalization strategies. Let $minRank(v,C_i)$ denote the minimal rank of node $v$ based on normalization strategy $C_i$ over all $\beta$-values and define $maxRank(v,C_i)$ accordingly. Then, $\Delta agg(v):=\max\{maxRank(v,C_i)-minRank(v,C_i)|1\leq i\leq 4\}$, i.e., the maximum over all individual maximal ranking differences within one normalization methods, over all normalization methods. While  $\Delta agg(Manchester)=5$, it is only $\Delta agg(Francisco)=2$ for Francisco. 	

Similarly, for any $\beta$-value, the sensitivity on the chosen normalization method can be determined as the maximal difference in ranking based on the different normalization methods. Let $maxRank(v,\beta)$ denote the maximal rank of $v$ based on any normalization method and let $minRank(v,\beta)$ be defined accordingly. For example, Francisco at  $\beta=20$ shows a maximal difference in the ranking positions of $7-2=5$. The overall sensitivity of a node on the chosen normalization strategy is then defined as $\Delta norm(v):=\max\{maxRank(v,\beta)-minRank(v,\beta)|\beta in \Gamma\}$, where $\Gamma$ is a set of different $\beta$-values. Both, Manchester and Francisco, have a $\Delta norm$ value of $5$. 

 Keeping this in mind, if the layer of Lufthansa is removed from the data set, there are $20$ common nodes (airports) among the remaining three layers representing Air Berlin, Easyjet, and Ryan airlines. The results of ranking for some nodes selected from the $20$ airports are shown in Figure.~\ref{Fig1}.b). Interestingly, Manchester airport's sensitivity has decreased a bit: $\Delta agg(Manchester)$ is now only $5$ within the larger data set and $\Delta norm(Manchester)$ is only $2$. Francisco, however, has now a $\Delta agg$ of $6$, and a $\Delta norm$ of $4$. An airport like Chania does not show any sensitivity to the normalization strategy, as all four curves fall on top of each other, and only a very small sensitivity against the aggregation strategy: $\Delta agg(Chania)=1$. Finally, Venice is much more sensitive against the aggregation strategy than the normalization strategy ($\Delta agg(Venice)= 10,\ \Delta norm(Venice)=3)$.

In general, looking at the $\Delta agg$ and $\Delta norm$ values of a node, it can be found that a node is in one of four categories: 
\begin{enumerate}
\item Sensitive only to the choice of the normalization strategy ($A0N+$), 
\item Sensitive only to the choice of the aggregation strategy ($A+N0$), 
\item Sensitive to both ($A+N+$), and 
\item Sensitive to none ($A0N0$).
\end{enumerate}
 
 \begin{figure}[!h]
 	\includegraphics[width=0.5\textwidth]{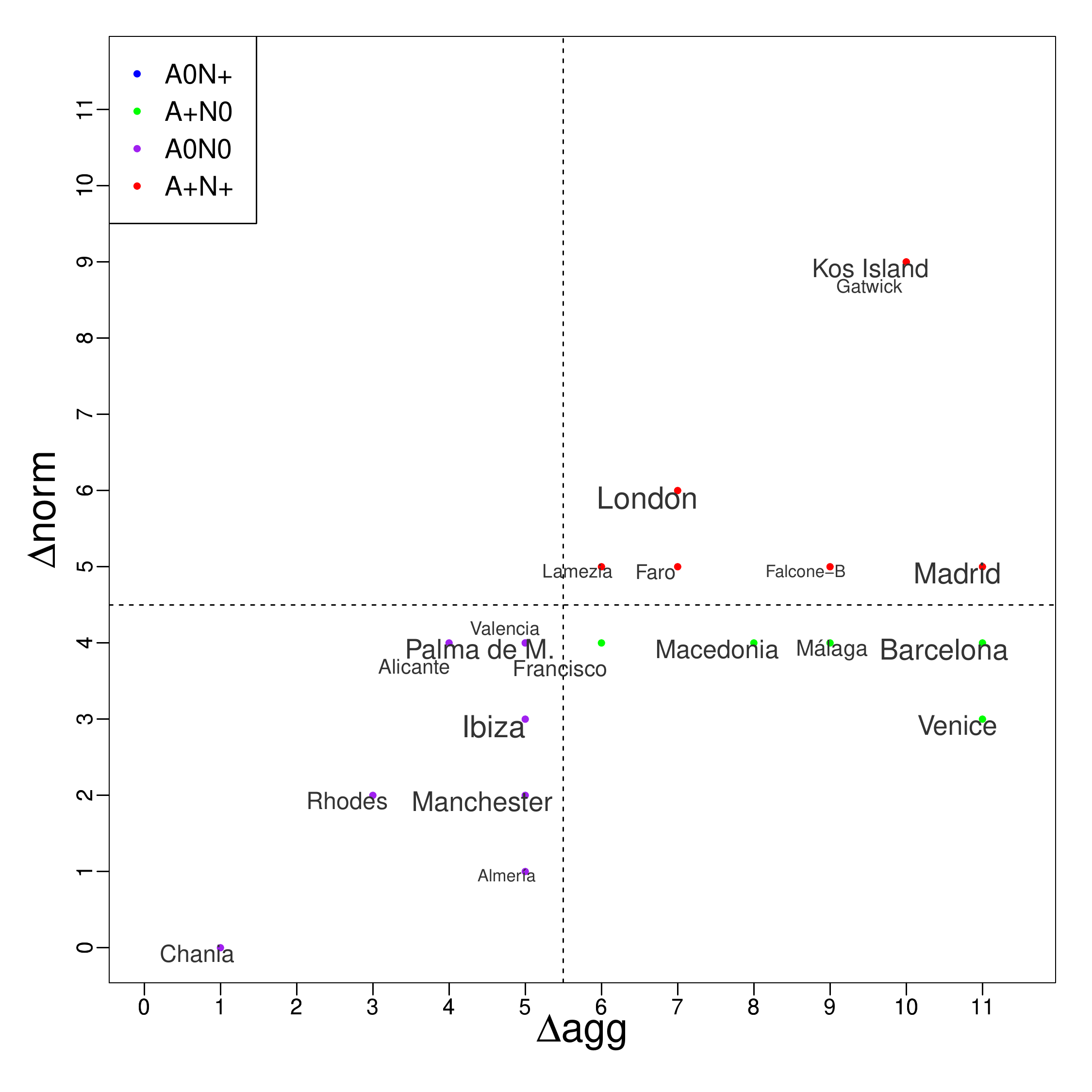}
 	\caption{European airlines network dataset. Sensitivity of the $20$ common airports with respect to the choice of aggregation strategy and normalization strategy, as quantified by $\Delta agg$ and $\Delta norm$, respectively. The four sections of plot respectively contain the group of nodes sensitive to only one choice ($A0N+$ or $A+N0$), those sensitive to none ($A0N0$), or both ($A+N+$). 
	\label{RD-air}}
 \end{figure}

Thus, Fig.~\ref{RD-air} shows a scatter plot of these two sensitivity values for each node in the European Airline data set. As expected, Chania as a very robustly ranked airport is in the bottom left of the plot, and two sensitive airports like Kos Island and Gatwick are in the top right of the plot ($\Delta agg = 10$, $\Delta norm=9$).  While for most airports the two sensitivity measures are correlated, some airports like Venice and Barcelona are more sensitive to the aggregation than to the normalization strategy. There is no airport with the opposite behavior. 
 
The airline data set is good as a small example which yields first, interesting insights. However, the visualization shown in Fig.~\ref{RD-air} can also be used for larger data sets to determine whether and how many nodes are sensitive to the choice of the normalization and aggregation strategy. 

\begin{figure}[]
	\includegraphics[width=0.5\textwidth]{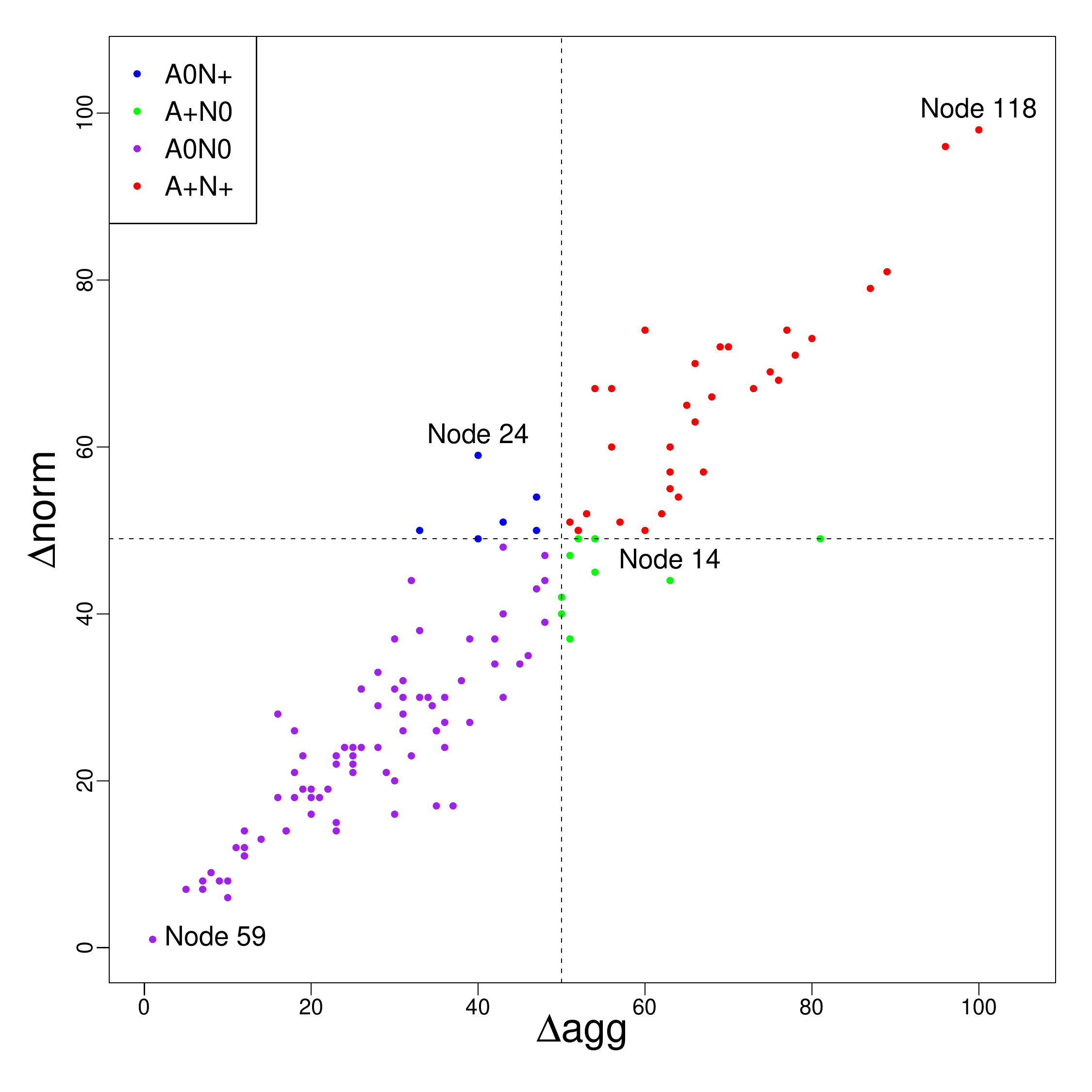}
	\caption{Twitter network dataset. Ranking difference of nodes with respect to the different aggregation strategies ($\Delta agg$) and the different normalization methods ($\Delta norm$). The four sections of plot respectively contain the group of nodes sensitive to only one choice ($A0N+$ or $A+N0$), those sensitive to none ($A0N0$), or both ($A+N+$).	\label{RD-Higgs} 
	}
\end{figure}

\begin{figure}
	\includegraphics[width=0.5\textwidth]{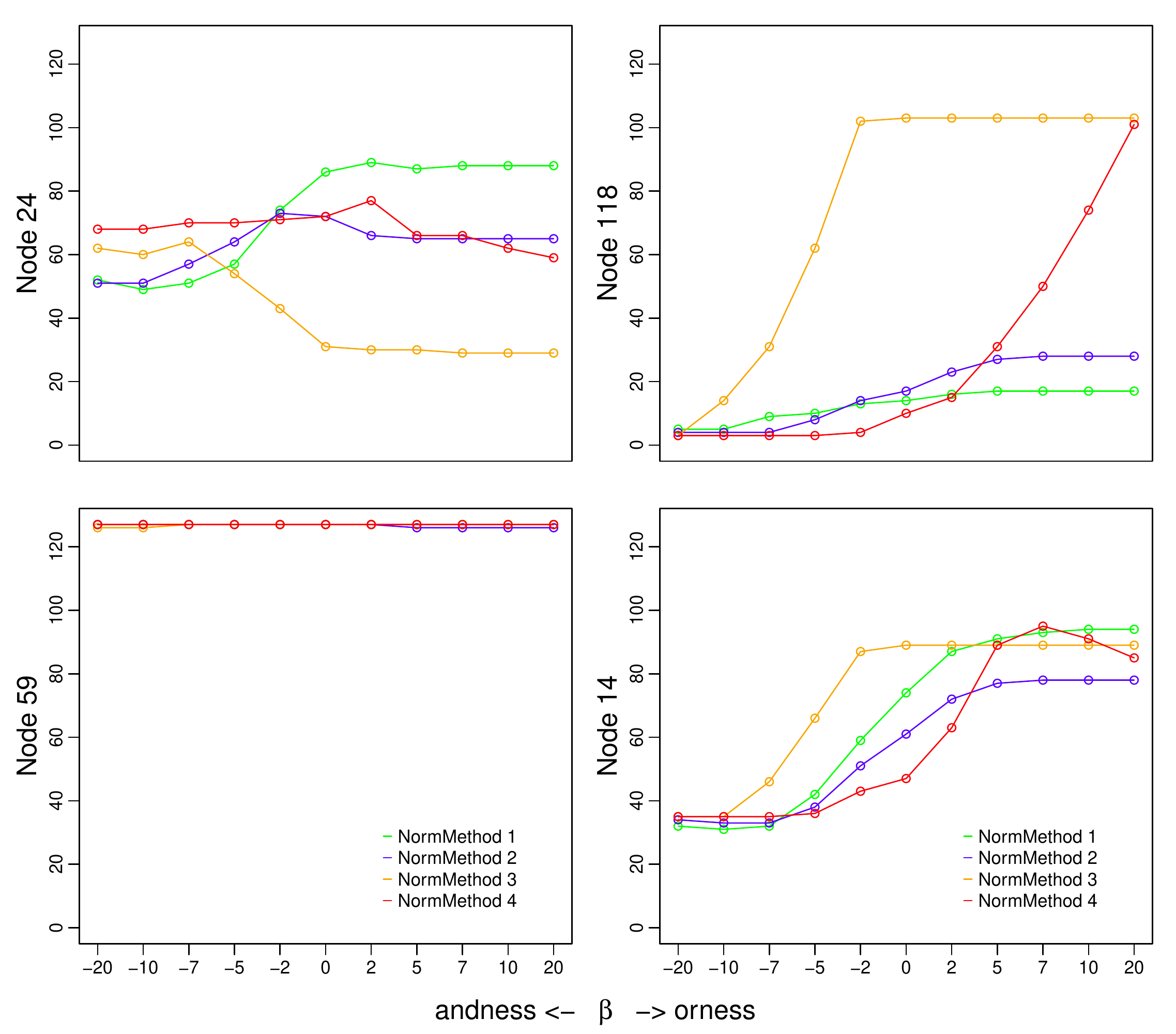}
	\caption{Twitter network data set. Rankings of the four nodes that are positioned in the four sections of $\Delta agg$ and $\Delta norm$ scatter plot are depicted here. Their rankings positions are obtained using different aggregation strategies (guided by $\beta$) from the four layers ({\it mentioning, reply, retweet}), and {\it social network layer}). The curves show the results obtained using the four normalization strategies. \label{Higgs}}
\end{figure}

Fig.~\ref{RD-Higgs} shows the corresponding plot for the Twitter data set. 
Figure~\ref{fig:3cumdist} b) already showed that a high percentage of nodes in the {\it social network} layer has small normalized degrees compared to the corresponding values in the other two layers ({\it retweet} and {\it reply}).  We discuss four nodes out of $127$ common nodes in all four layers; these are marked in Fig.~\ref{RD-Higgs}. The four nodes represent each of the four categories denoted above. Fig.~\ref{Higgs} shows their ranking curves. 

A node in the most bottom left position in Figure~\ref{RD-Higgs}---where the location of the robustly ranked group is ($A0N0$)---is node 59. Its absolute degrees in the different layers are $[141,29,101,33664]$. Its ranking curves shown in Figure~\ref{Higgs} show that the ranking of node 59 is almost stable using any combination of the four normalization methods and different aggregation strategies (the different $\beta$ values). Comparing its degrees with the maximal degrees shown in Table~\ref{tab:threetables} b), it can be found that in the three layers of {\it mentioning}, {\it reply} and in the {\it social network}, node 59 obtains the maximum total degree among ($|V^*|=127$) common nodes. It is thus always placed among the top five nodes and one of the most stable ranked nodes. In the most top right position, node 118 is highlighted with absolute degree values of $[6, 2, 2, 1396]$; the ranking curves of node 118 are shown in Figure~\ref{Higgs}. This node has the highest sensitivity to both, the choice of the aggregation and normalization strategy. It can be seen that its ranking based on NormMethod 3 varies from $3$ to $103$ --from bottom 3 to top 24 nodes-- using different aggregation strategies---the same applies for NormMethod 4. Interestingly, these two methods also provide the biggest difference at any given $\beta$-value, namely at $\beta=0$. Thus, node 118 has the highest sensitivity to both choices. 

Node 24 is more sensitive against the choice of the normalization than against the choice of the aggregation strategy with $\Delta norm=60$ and $\Delta norm =40$. It has absolute degree values of $[35,4,22,188]$. As can be seen in its ranking curves in Figure~\ref{Higgs}, the different normalization methods show three different kinds of behavior: NormMethod 1 has a downward trend, NormMethod 3 shows an upward trend while NormMethod 3 and 4 produce quite stable rankings. Finally, node 14 is located in the group that is rather sensitive to the aggregation than to the normalization ($\Delta agg=65$, $\Delta norm=46$).


\begin{figure}
	\includegraphics[width=0.5\textwidth]{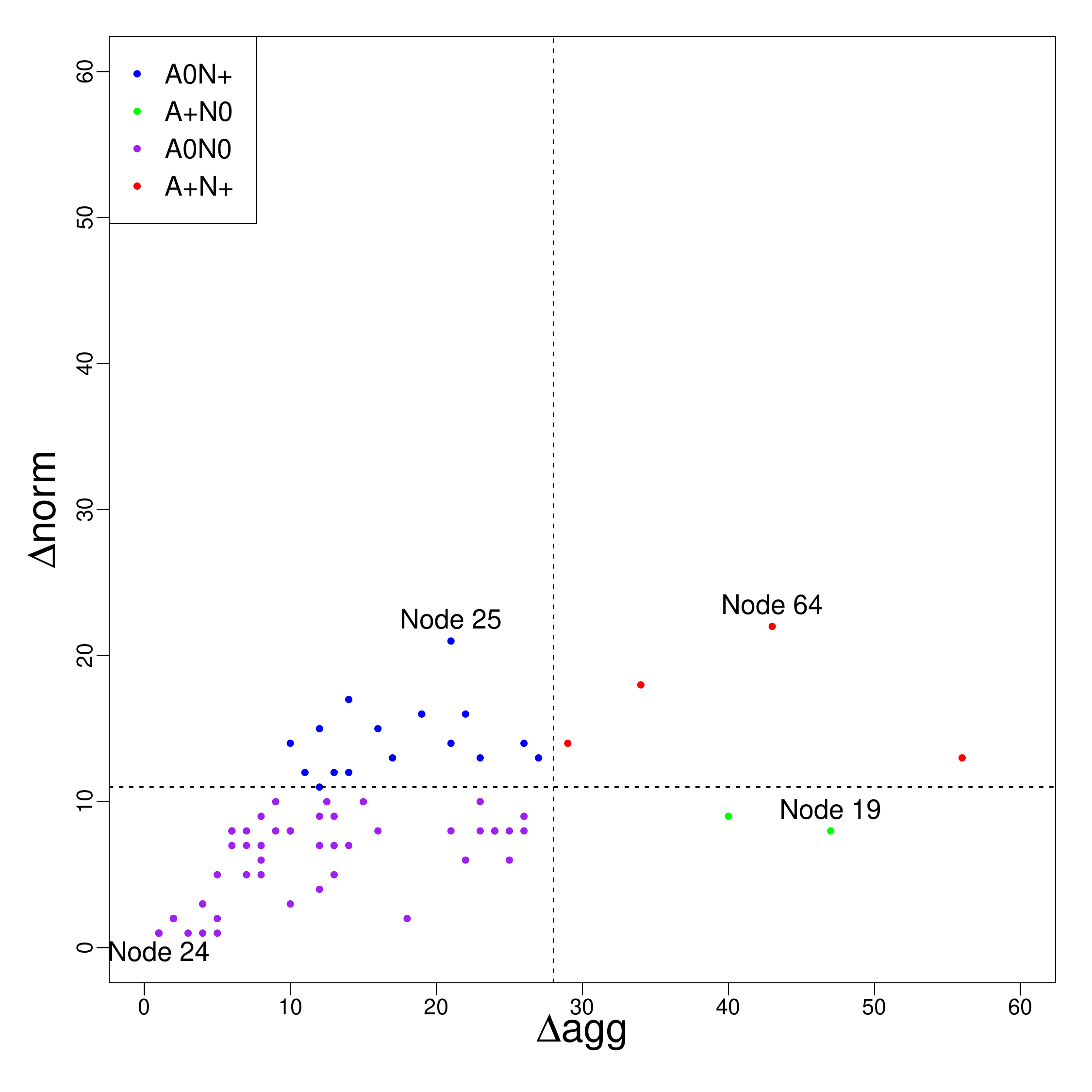}
	\caption{Law firm dataset.  Scatter plot of the $\Delta norm$ and $\Delta agg$ values of each node. The four sections of plot respectively contain the group of nodes sensitive to only one choice ($A0N+$ or $A+N0$), those sensitive to none ($A0N0$), or both ($A+N+$).
		 \label{RD-law}}
\end{figure}

\begin{figure}
	\includegraphics[width=0.5\textwidth]{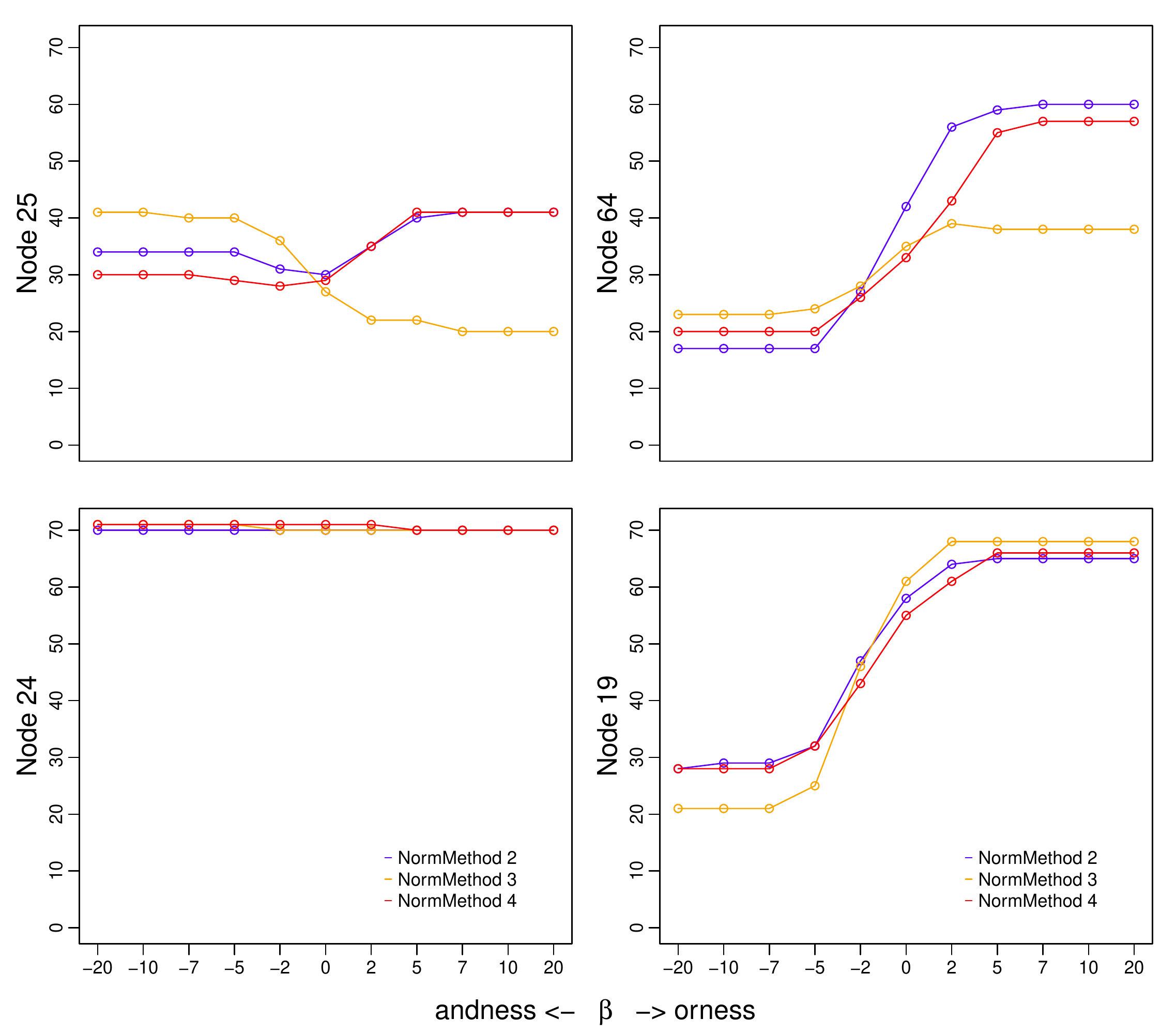}
	\caption{Law firm dataset. Rankings of the four nodes that are positioned in the four sections of $\Delta agg$ vs $\Delta norm$ scatter plot are shown here. The ranking positions obtained using the different aggregation strategies (using the $\beta$ parameter) for the aggregation of three layers of seeking advice, coworking, and friendship. The three curves show three different normalization methods (NormMethod 1 and NormMethod 2 give the same results in this dataset). \label{lawfirm}}
\end{figure}

However, not all multiplex networks show a large degree and variety of the sensitivity of their nodes. In the third dataset, the law firm dataset, a large fraction of the nodes are stable with respect to all combinations of aggregation and normalization strategies. Only a few nodes can be seen in the right side of Figure~\ref{RD-law}, i.e., six out of $71$ attorneys, all shared by all layers. However, still some interesting cases can be distinguished among the nodes. The ranking curves of four nodes $25,64,24$ and node $19$ selected from each group are depicted in Figure~\ref{lawfirm}. Node 24 is the most stable node. Looking at its ranking curves in Figure.~\ref{lawfirm}, it reveals that all of the curves are almost on top of each other. Node 64 instead, has the highest sensitivity to both modeling decisions with the maximum value of $\Delta norm=22$ among the nodes. Node 19 has the second highest value of $\Delta agg=47$ among the nodes. This is because using the different aggregation strategies, e.g., the maximum, minimum or the average over its three values, node 19 obtains more conflicting rankings. Note that the most sensitive node in the group of $A+N+$ has a $\Delta agg$ of $56$(!) in the most bottom right side of Figure~\ref{RD-law}. 
Node 25 has the second highest value of $\Delta norm=21$ and thus is located in the group of $A0N+$. Looking at its rankings in the three curves, it reveals that NormMethod 2 and NormMethod 4 produce similar rankings but NormMethod 3 results in a rather different ranking compared to the other two.


In a general view, as can be observed in Figure~\ref{fig:3cumdist}c), the cumulative distributions of the normalized degree values are quite similar in the three layers of seeking advice, coworking, and friendships in law firm dataset 
and this results in a low difference among the normalization methods as illustrated in Figure~\ref{RD-law}.

\section{\label{summary}Conclusion}
The intuitive explorations of nodes' degree centrality in this paper shows that even seemingly simple and inconsequential modeling decisions, such as normalization and aggregation strategies, lead to very different ranking positions of the nodes in multiplex networks. In addition, the visual method introduced in this paper allows to categorize nodes by their sensitivity to either different aggregation or normalization strategies, to both, or to none. The small sized dataset of an airline network is the first promising example which results in very interesting insights. The large sized dataset of a tweet network shows even more interesting cases which have a very high sensitivity to the choice of both aggregation and normalization strategy. A further interesting case is found in the medium sized dataset of a law firm: it shows a high sensitivity to the choice of the aggregation strategy. The experimental results emphasize the need of documenting all kinds of preprocessing steps to make the resulting analysis reproducible and its interpretation analyzable.

 In the future work, we will analyze the influence of different modeling decisions on the nodes' rankings in multiplex networks with respect to the centrality indices of \textit{betweenness} and \textit{closeness}. 
\bibliographystyle{IEEEtran}
\bibliography{sudetavassoli}

\end{document}